\documentclass[runningheads]{llncs}

\usepackage{amsfonts}
\usepackage{amssymb}
\usepackage{graphicx}
\usepackage[utf8x]{inputenc}
\usepackage[ruled]{algorithm2e}
\usepackage{amsmath}

\begin{document}

\title{From computational ethics to morality: how decision-making algorithms can help us understand the emergence of moral principles, the existence of an optimal behaviour and our ability to discover it}
\titlerunning{From computational ethics to morality}
\author{Eduardo C. Garrido-Merchán, Sara Lumbreras-Sancho}
\date{March 2023}

\institute{Universidad Pontificia Comillas, Madrid, Spain \\
Universidad Pontificia Comillas - IIT, Madrid, Spain \\
\email{ecgarrido@icade.comillas.edu, slumbreras@comillas.edu}}

\maketitle 

\abstract{This paper adds to the efforts of evolutionary ethics to naturalize morality by providing specific insights derived from a computational ethics view. We propose a stylized model of human decision-making, which is based on Reinforcement Learning, one of the most successful paradigms in Artificial Intelligence.
After the main concepts related to Reinforcement Learning have been presented, some particularly useful parallels are drawn that can illuminate evolutionary accounts of ethics.
Specifically, we investigate the existence of an optimal policy (or, as we will refer to, objective ethical principles) given the conditions of an agent. In addition, we will show how this policy is learnable by means of trial and error, supporting our hypotheses on two well-known theorems in the context of Reinforcement Learning. We conclude by discussing how the proposed framework can be enlarged to study other potentially interesting areas of human behavior from a formalizable perspective.}


\section{Introduction}
In recent times, evolutionary ethics has attempted to explain morality using naturalistic principles that rely on empirical observations of the natural world, including the evolutionary history of humans. It is a multidisciplinary field that draws upon knowledge from various fields, such as biology, psychology, anthropology, and philosophy. The origins of this approach can be traced back to Darwin's theory of evolution by natural selection, which suggested that humans are not fundamentally different from other animals and instead are part of a larger continuum of life that has evolved over millions of years. This view challenged many traditional moral beliefs, including the notion that humans have a unique status or purpose in the universe. Instead, it implied that our moral sense is a result of evolution, shaped by natural selection to encourage cooperation and social harmony within groups. The efforts to understand the emergence of morality from the principles of natural selection have crystallized in the field of evolutionary ethics \cite{allhoff2003evolutionary}.

Similarly, natural selection has been used to explain the evolution of culture. Cultural evolution is based on the idea that cultural traits, such as ideas, beliefs, and behaviors, can be passed down from one generation to the next and can evolve over time through various mechanisms \cite{distin2011cultural}. These mechanisms include innovation, diffusion, and selection, which can be put in parallel with the biological mechanisms of mutation, crossover and selection. Cultural evolution has become an important area of research as it can help explain the development of complex societies, the spread of religions and languages, and the evolution of technology.

These ideas have been further refined by the introduction of multilevel selection \cite{young2011collective}, which explains that not only the individual traits are selected for, but also the group ones. In fact, several levels are operating at the same time, for instance, with individuals, families, companies, social classes or countries. The concept of multilevel selection pertains to the evolutionary influence of group dynamics as opposed to individual adaptive pressures. Interestingly, the concept of multilevel selection has been increasingly applied to human behavior across different areas such as development and economic dynamics. 

Interestingly, some authors \cite{young2011collective} have speculated that the different levels of multilevel selection (such as individual, kin, and larger groups) could have worked together to help evolve the five proposed Neo-Piagetian stages in human cognitive development, from reflexive and sensorimotor to abstract and collective intelligence.

The perspectives that attempt to naturalize Ethics are complementary to the existing philosophical views on morality. While the debates in this field are vast and intricate, we will simplify them to the contrast between ethical realism and relativism. Ethical realism is a philosophical stance that asserts the existence of moral truths that are independent of subjective human opinions or beliefs. It proposes that ethical statements can be true or false and are founded on objective aspects of the world. In essence, ethical realism maintains that moral values and principles are not merely human conventions or expressions of personal preferences, but rather stem from an objective moral reality that exists regardless of individual perspectives. This perspective differs from moral relativism, which refutes the existence of objective moral truths and claims that moral values and principles are contingent on cultural, historical, or personal contexts. Some argue that only relativism can acknowledge the diversity of human perspectives and cultural values \cite{relativism2008stanford}, but others argue that it can lead to a dangerous moral relativism and nihilism \cite{mchoskey1999relativism}. The supporters of ethical realism have paved the ground to the elicitation of a global ethic which seeks to identify shared moral values that can be universally applied across cultures and religions \cite{kung1998global}. In particular, this approach emphasizes the importance of recognizing our common humanity and working towards a shared vision of justice and peace. While there are certainly challenges to implementing a global ethic, many argue that it offers a powerful framework for addressing some of the most pressing ethical dilemmas facing our world today \cite{singer2011expanding}. Nonetheless, some scholars still challenge the possibility of a global ethic, arguing that ethical norms are inevitably shaped by cultural context and cannot be universally applied \cite{appiah2017cosmopolitanisms}.

Interestingly, it is possible to explore this dilemma from a mathematical point of view, extracting moral insights from the implications of ethical models built consistently with the naturalization efforts from multilevel selection and evolutionary ethics.

In particular, we propose to use reinforcement learning, one of the most successful paradigms in Artificial Intelligence, to build a metaphor of how human decision making is performed. Reinforcement Learning is a branch of machine learning concerned with the development of algorithms that enable agents, that in this context would be people, to learn to make decisions based on feedback received from the environment, that in this context would be the world that we perceive. In reinforcement learning, an agent interacts with an environment and learns to select actions that maximize a reward signal \cite{sutton2018reinforcement,kaelbling1996reinforcement}. This process is guided by a policy, which maps states to actions (that is, a policy is a guide which shows what is the preferred action for any given state). In our metaphor for human decision-making, policies will be used to represent ethical guidelines, as they are the mathematical expression of "if under situation s, the agent should do action a".

Accepting a set of assumptions that will be further discussed in this paper, the environment, our world, could be modeled as a Markov decision process (MDP) \cite{puterman1990markov}, which is a mathematical framework for modeling decision-making in situations where outcomes are partly random and partly under the control of a decision maker, the people. Many real-world problems, such as robotics control \cite{kober2013reinforcement}, game playing \cite{silver2017mastering}, and autonomous driving \cite{aradi2020survey}, have been formulated as MDPs, and we can argue that under some reasonable simplifications it is possible to model with a MDP the environment faced by an individual and her set of possible actions.

It should be noted that our model of ethical action by a MDP and using Reinforcement Learning does not coincide with the main perspectives on the emerging field of Computational Ethics. Computational Ethics is concerned with the creation and utilization of technological means to formalize and appraise ethical decision-making. This entails the application of ethical decision-making processes, the deployment of mathematical and computational models to represent moral tenets, and the use of simulations and experiments to assess moral behavior. The overarching objective is to establish a structure for integrating ethical considerations into computer algorithms and other automated decision-making systems. In our work, we do use mathematical and computational models to represent moral tenets. However, our intention is not to integrate ethical considerations into automated decision-making systems but rather deepen our understanding of how ethical judgments are passed and ethical values derived.

In particular, our contribution focuses on building a model for decision-making based on reinforcement learning and drawing from the mathematical properties of reinforcement learning to the properties of ethical reasoning. The concrete insights are derived from three well-studied theorems of the reinforcement learning field. The first one shows that, given the definition of an environment and a set of possible actions, an optimal decision policy always exists. The second theorem proves that this optimal policy can be learned by agents. The last one shows that learning can happen by a simple trial-and-error process.

The organization of the paper is as follows. First, we review the most important concepts related to reinforcement learning. Then, we build our metaphor of human decision-making based on the formal model of reinforcement learning. Immediately later, the three reinforcement learning theorems are presented, and we analyze their implications with respect to moral reasoning. Finally, we conclude  the paper with a discussion and conclusions and further work sections.

\section{Reinforcement learning as decision-making}
Reinforcement learning (RL) is a subfield of machine learning that focuses on training agents to make decisions by interacting with an environment. It is fundamentally rooted in the Markov Decision Process (MDP) framework, which is defined by a tuple $(\mathcal{S}, \mathcal{A}, P, R, \gamma)$, where $\mathcal{S}$ is the state space, $\mathcal{A}$ is the action space, $P$ represents the state transition probabilities, $R$ is the reward function, and $\gamma$ is the discount factor. In particular, the goal of an RL agent is to learn an optimal policy $\pi: S \to \mathcal{A}$, which is a mapping from states to actions that maximizes the expected cumulative discounted reward. The value function $V_{\pi}(s)$ is defined as the expected cumulative discounted reward the agent receives when starting in state s and following policy $\pi$, formally expressed as $V_{\pi}(s) = \mathbb{E}[\sum_{t=0}^{\infty} \gamma^t R(s_t, a_t) | s_0 = s, \pi]$, where $R(s_t, a_t)$ is the immediate reward at time step $t$. If we assume that the agent is the brain and the environment is the world, the problem can be modeled as in Figure \ref{exp1}, where the agent would have a model of the world as in Figure \ref{exp2}.

\begin{figure}[h]
    \centering
    \includegraphics[width = 0.6\textwidth]{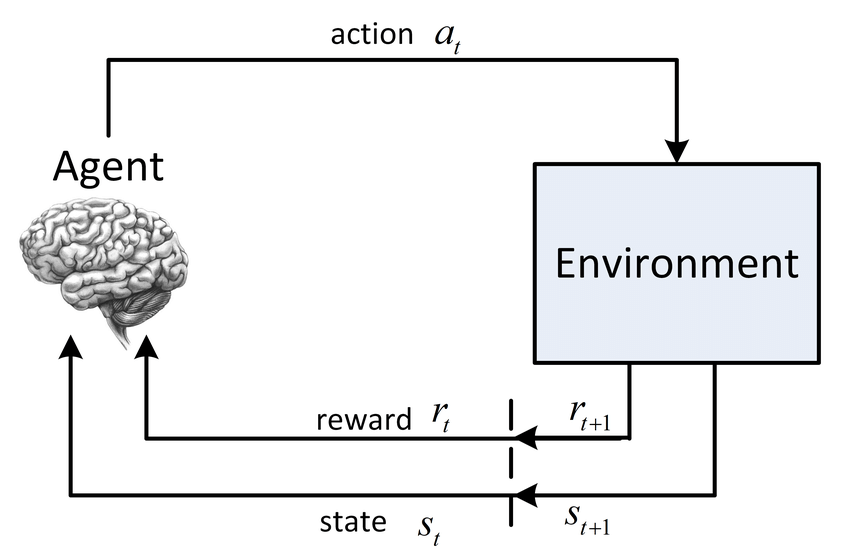}
\caption{The relationship between the environment an agents in a Markov Decision Process, where the human brain is the agent.}
    \label{fig:exp1}
\end{figure}

Critically, Q-learning is a widely-used, model-free (this means that it is free from any assumptions about the working of the outside world) reinforcement learning algorithm that aims to learn the optimal action-value function $Q\star(s, a)$, which represents the expected cumulative discounted reward for taking action a in state s and following the optimal policy thereafter. The Bellman optimality equation characterizes $Q\star$ as follows: $Q\star(s, a) = \mathbb{E}[R(s, a) + \gamma max_{a'} Q\star(s', a')]$, where $s'$ is the successor state after taking action a in state $s$. Q-learning iteratively updates the estimated action-value function, $Q(s, a)$, based on observed transitions $(s, a, r, s')$ using the update rule: $Q(s, a) = Q(s, a) + \alpha [r + \gamma max_{a'} Q(s', a') - Q(s, a)]$, where $\alpha$ is the learning rate. The Q-learning algorithm guarantees convergence to the optimal action-value function $Q\star$ under certain conditions, such as using a sufficiently small learning rate and exploring the state-action space infinitely often.

\begin{figure}[h]
    \centering
    \includegraphics[width = 0.4\textwidth]{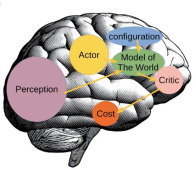}
\caption{Graphical description of a potential deep reinforcement learning world model of the human brain \cite{matsuo2022deep}.}
    \label{fig:exp2}
\end{figure}

Even if the processes that underlie human decision-making could be much more complex than Q-learning (for once, human beings are able to use reasoning to anticipate the consequences of their actions and not rely solely on empirical evidence), we will hypothesize that the main elements that are present in reinforcement learning are also present in human decision making: an agent capable of actions, and environment capable of states and the existence of a reward for the agent that is dependent of the state and that is maximized.

\section{The crucial role of the reward function}

Although the model, as described above, is overly simplistic, it should be noted by the reader that it is sufficiently flexible to account for most of the relevant features of ethical reasoning. Many of them are accounted for in this section.

First, the definition of the reward function is key. 
A very useful attempt at understanding the reward function in human behavior is dopamine reward hypothesis. Concretely,  

\begin{center}
\textit{"That all of what we mean by goals and purposes can be well thought of as the maximization of the expected value of the cumulative sum of a received scalar signal, called reward."}
\end{center}

In particular, the dopamine reward hypothesis posits that dopamine, a neurotransmitter, plays a crucial role in mediating the experience of reward and reinforcement, as well as driving motivated behaviors in animals and humans \cite{wise2004dopamine}. This hypothesis is supported by a wealth of experimental evidence, including the observation that dopamine neurons in the ventral tegmental area (VTA) and substantia nigra (SN) project to regions involved in processing rewards, such as the nucleus accumbens (NAc) and the prefrontal cortex (PFC) \cite{ikemoto2007dopamine}. Electrophysiological studies have demonstrated that dopamine neurons exhibit phasic activity in response to both unexpected rewards and reward-predicting cues, consistent with a role in encoding reward prediction errors \cite{schultz2016dopamine}. The dopamine reward hypothesis has also been used to explain maladaptive behaviors, such as addiction, in which increased dopaminergic activity has been linked to the reinforcing effects of drugs of abuse \cite{volkow2015brain}. In particular, the reward $r$ being lower than $\infty$ is justified just because the brain does not deal with $\infty$ dopamine.

Even if the reward function was substantially more subtle, for instance, including more neurotransmitters than dopamine, or a more complex assessment of action consequences, the main implications from using the reinforcement learning metaphor would still be valid, as they do not depend on any particular definition of reward or model of the environment.

Studying the reward function further, it is necessary to consider whether it is defined in an objective vs. individual manner. If the reward function is defined in an objective manner, using the formal language of mathematics, we would have a framework consistent with consequentialism  \cite{bentham1996collected}. Under a consequentialist view, the morally right action is the one that produces the best overall outcome or consequence. This means that the end justifies the means, and the focus is on maximizing good consequences while minimizing bad consequences.

Now, the reward function can be defined in a global manner by taking into account the full humanity. This would lead to utilitarianism, which holds that the morally right action is the one that maximizes overall happiness or pleasure. On the contrary, if the reward function is defined based only on parameters that pertain to the individual, we can formalize ethical egoism, which argues that individuals should act in their own self-interest to maximize their own well-being.

Crucially, it is possible to define a hierarchy of objectives that can cover from the individual to the narrow group, the larger group, and the whole of humanity. This model echoes multi-level selection and would be able to capture the layered dynamics that underlie human behavior.

In addition, if the reward function is expressed in a manner that includes some purely individual aspects, it would be possible to include some of the nuances and complexities of human behavior. For instance, personality traits or specific tastes could be applied as a filter to calculate the reward corresponding to a state, in the same way that utility functions are applied in economics to represent the satisfaction or pleasure that consumers receive for consuming a good or service.

Now, the main theorems that apply to reinforcement learning will be presented and the conclusions for ethical behavior will be explained.

\section{The existence of an optimal policy} 

Before presenting the first of the theorems in this paper, it is necessary to explain what an infinite horizon discounted Markov Decision Process (MDP) is and why reality experienced by a human could be modelled approximately by such a MDP. Let $\mathcal{A}$ be the action space of a human being, $\mathcal{S}$ the observation space captured by the perception of the human being and $r$ the reward given to a human being as a result of performing action $a \in \mathcal{A}$ as a response to $s \in \mathcal{S}$. We would have to justify that the number of possible actions $\mathcal{A}$ done by a human being is lower than $\infty$ and the number of observed situations is also lower than $\infty$. 

\begin{theorem}
Assume $\vert \mathcal{A} \vert < \infty, \vert \mathcal{S} \vert < \infty$ and $\vert r \vert < \infty$ with probability one. For any infinite horizon discounted MDP, there always exists a deterministic stationary policy that is optimal.
\end{theorem}

This is theorem 6.2.7 in \cite{puterman2014markov}. Concretely, a deterministic stationary policy $\pi : \mathcal{S} \to \mathcal{P}(\mathcal{A})$ is a mapping from the state space $\mathcal{S}$ to the action space $\mathcal{A}$, where $a \in \mathcal{A}$ is the action selected by the policy in state $s$, that assigns a fixed probability distribution over actions for each state. This result has important implications for ethics as if the Markov Decision Process model of the world
perceived by agents assumption is valid, it guarantees that a simple policy representation with a fixed mapping from states to actions can achieve optimal performance in the ethics class of problems.  

The philosophical assumption that the world can be modeled as a Markov Decision Process (MDP) perceived by human agents suggests that the complexity of human experience can be reduced to a series of states, actions, and rewards, with each decision being based on the current state and the probabilities of transitioning to future states. This perspective builds upon the idea that humans are rational decision-makers, striving to optimize their actions in order to maximize some objective function, often represented as the expected cumulative reward over time, as explained in the previous section. Proponents of this view argue that the MDP framework is a useful abstraction for understanding and predicting human behavior across diverse domains, such as economics, psychology, and artificial intelligence \cite{daw2014algorithmic}. However, critics of this philosophical assumption contend that it oversimplifies the intricacies of human cognition, emotion, and social interaction, neglecting the influence of context, memory, and individual differences in decision-making processes \cite{sloman2005problem}. Despite these criticisms, the MDP framework has provided valuable insights into human behavior and decision-making, while also inspiring the development of computational models and algorithms that mimic human-like intelligence \cite{lake2017building}.

\begin{figure}[h]
    \centering
    \includegraphics[width = 0.8\textwidth]{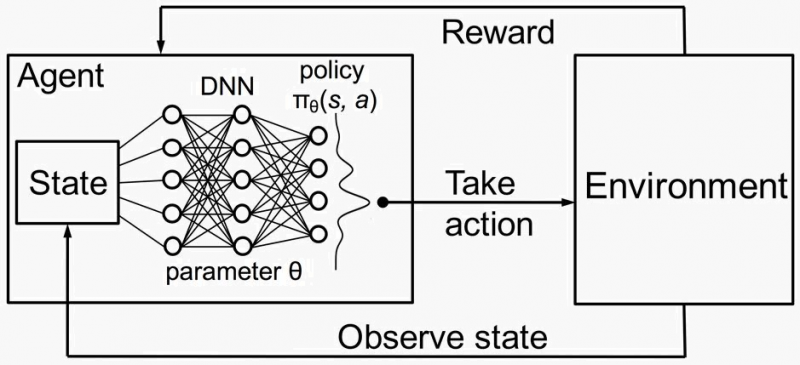}
\caption{Deep reinforcement learning modelization of an agent (human being) whose external behavior is parametrized via a deep neural network that learns the ethical optimal policy.}
    \label{fig:drl}
\end{figure}

We could argue that the world contains a nearly infinite number of states $s \in \mathcal{S}$ and so the reinforcement learning model optimizing decision-making would need a colossal number of parameters $\theta$ to infer the optimal policy $\pi$, that is, the optimal action $a$ conditioned to every state $s$. However, it could also be reasonable to assume that the description of the world as understood by agents is not complete but only partial and is adapted while they interact with it via their actions. A schematic description of this is shown in Figure \ref{fig:drl}.

This means that, given a definition of the possible states of the environment, the spectrum of actions and the reward function, there is mathematically an optimal policy that links states to actions. We will take this to mean the following: given the specifications of an ethical problem, it is always possible to define an ethical (that is, the optimal given the reward) vs. an unethical one. This finding would explain relativism (individual or cultural) just by differences in the reward function (that is, whether some things are more valued than others by different individuals or in different cultures).

This also means that, if we formalize survival as the reward function (which is consistent with the point of view of evolutionary ethics), it would be possible to derive some objective policy as optimal for the situation. This becomes even clear when multilevel selection is considered and the uppermost 'humanity' level is taken into account: there is an objective policy that maximizes the possibilities of human survival.

\section{Computability and learnability}
Now that we have shown how under the described assumptions, an objective ethic exists, we will show how, given a set of conditions, this policy is computable, that is, can be calculated.

First, we need to describe the concepts of bounded reward, adaptive learning rate $\beta_n$ and Q-function $Q^{(n)}$. In particular, a reward is said to be bounded if there exists a constant $M > 0$ such that the absolute value of any reward is limited, i.e., $|R(s, a, s')| \leq M$ for all $s, a, s'$. Bounded rewards help ensure convergence of learning algorithms and stability in long-term value estimation. The adaptive learning rate $\beta_n$ is a dynamically updated parameter that controls the step size in the learning process. Lastly, the Q-function, denoted as $Q^{(n)}(s, a)$, represents the expected cumulative discounted reward when taking action $a$ in state $s$ and following a policy $\pi^{(n)}$ thereafter. Now, we can illustrate the following theorem:

\begin{theorem}
    Assume $\vert \mathcal{A} \vert < \infty$ and $\vert \mathcal{S} \vert < \infty$. Let $n^i(s,a)$ be the index of the i-th time that the action a is used in state s. Let $R < \infty$ be a constant. Given bounded rewards $\vert r \vert \leq R$, learning rate $0 \leq \beta_n < 1$ and:

    \begin{align}
        \sum_{i=1}^{\infty} \beta_{n^i(s,a)} = \infty , \sum_{i=1}^{\infty}(\beta_{n^i(s,a)})^2 < \infty, \quad \forall s,a,  
    \end{align}

    then $Q^{(n)}(s,a) \to Q^*(s,a)$ as $n \to \infty, \forall s,a$ with probability 1.
\end{theorem}

This theorem was shown on Watkins and Dayan (1992) \cite{watkins1992q}. It basically says that a series of conditions exist that guarantee the convergence to a true value function $Q^\star$ that would be the optimal global objective ethic in the MDP that represents our world. In particular, this optimal policy does not only exist but is also feasible and learnable by analyzing the gradient of the objective function $J(\theta)$.

This means that the optimal policy for human beings (which we will call objective ethic) exists and can be practiced via the action space that has been defined.

In addition, as the last theorem will show, this policy is learnable.

\begin{theorem}
(Policy Gradient Theorem \cite{sutton1999policy}) Assume that $\pi(s,a;\theta)$ is differentiable with respect to $\theta$ and that there exists $\mu^{\pi_\theta}$, the stationary distribution of the dynamics under policy $\pi_\theta$, which is independent of the initial state $s_0$. Then the policy gradient is:

\begin{align}
    \nabla_\theta J(\theta) = \mathbb{E}_{s \sim \mu^{\pi_\theta}, a \sim \pi_{\theta}}(\nabla_\theta ln \pi(s,a;\theta)Q^{\pi_\theta}(s,a)).
\end{align}
\end{theorem}

The previous theorems imply that, under the assumptions explained in previous sections, an objective ethic exists, it can be followed by human beings and we can learn it through experience.

\section{Conclusions and further work}
In this work, we have built a formalization of human decision-making that builds on the naturalization attempts of evolutionary ethics and multilevel selection. Our formalization is based on reinforcement learning, one of the most successful paradigms in Artificial Intelligence. Then, three key theorems describing the properties of reinforcement learning systems have been presented and applied to the case of human decision-making.
They imply that, given a definition of the reward function, states of the world, and possible actions, it is possible to define an optimal policy that maps states to actions and that will be used to represent an objective ethics.
Then, we have shown that this objective ethics is feasible and learnable by the agents by experience.
Further work should analyze how the specificities of the reward function, including individual and group features, can give rise to specific phenomena that have been observed in the context of evolutionary ethics.
Indeed, the main point where reflection is needed is the reward function. In the development of computer systems, a programmer creates a mathematical function to express their preferences. We can define such functions externally, but it is easy to imagine that one of this functions could be intrinsic to living organisms. Survival and reproduction are commonly considered as the ultimate goal for a living being, with intermediate goals chosen by natural selection that support survival. This means that not only levels above the individual exist, but also within the individual, making matters even more complex. Homeostasis is one such process, which includes physical integrity and feeding when hungry. 

Reinforcement learning provides an intuitive understanding of howthe impact of intermediate goals on the final goals can be derived. For example, group membership can be critical for survival in many animal species, and the importance of belonging to a group can be ingrained genetically as strongly as the fight for physical integrity. However, certain behaviors that violate social norms must typically be learned through experience. Humans can contravene social norms or make decisions that endanger their survival if they deem it necessary.

Our reinforcement learning framework, therefore, can provide with a quantitative support to understand the objective basis for ethical reasoning and a possible mechanism for ethical learning.

\bibliography{main}

\begin{thebibliography}{10}

\bibitem{allhoff2003evolutionary}
{\sc Allhoff, F.}
\newblock Evolutionary ethics from darwin to moore.
\newblock {\em History and philosophy of the life sciences\/} (2003), 51--79.

\bibitem{appiah2017cosmopolitanisms}
{\sc Appiah, K.~A.}
\newblock {\em Cosmopolitanisms}.
\newblock NYU Press, 2017.

\bibitem{aradi2020survey}
{\sc Aradi, S.}
\newblock Survey of deep reinforcement learning for motion planning of
  autonomous vehicles.
\newblock {\em IEEE Transactions on Intelligent Transportation Systems 23}, 2
  (2020), 740--759.

\bibitem{bentham1996collected}
{\sc Bentham, J.}
\newblock {\em The collected works of Jeremy Bentham: An introduction to the
  principles of morals and legislation}.
\newblock Clarendon Press, 1996.

\bibitem{daw2014algorithmic}
{\sc Daw, N.~D., and Dayan, P.}
\newblock The algorithmic anatomy of model-based evaluation.
\newblock {\em Philosophical Transactions of the Royal Society B: Biological
  Sciences 369}, 1655 (2014), 20130478.

\bibitem{distin2011cultural}
{\sc Distin, K.}
\newblock {\em Cultural evolution}.
\newblock Cambridge University Press, 2011.

\bibitem{ikemoto2007dopamine}
{\sc Ikemoto, S.}
\newblock Dopamine reward circuitry: two projection systems from the ventral
  midbrain to the nucleus accumbens--olfactory tubercle complex.
\newblock {\em Brain research reviews 56}, 1 (2007), 27--78.

\bibitem{kaelbling1996reinforcement}
{\sc Kaelbling, L.~P., Littman, M.~L., and Moore, A.~W.}
\newblock Reinforcement learning: A survey.
\newblock {\em Journal of artificial intelligence research 4\/} (1996),
  237--285.

\bibitem{kober2013reinforcement}
{\sc Kober, J., Bagnell, J.~A., and Peters, J.}
\newblock Reinforcement learning in robotics: A survey.
\newblock {\em The International Journal of Robotics Research 32}, 11 (2013),
  1238--1274.

\bibitem{kung1998global}
{\sc K{\"u}ng, H.}
\newblock {\em A global ethic for global politics and economics}.
\newblock Oxford University Press on Demand, 1998.

\bibitem{lake2017building}
{\sc Lake, B.~M., Ullman, T.~D., Tenenbaum, J.~B., and Gershman, S.~J.}
\newblock Building machines that learn and think like people.
\newblock {\em Behavioral and brain sciences 40\/} (2017), e253.

\bibitem{matsuo2022deep}
{\sc Matsuo, Y., LeCun, Y., Sahani, M., Precup, D., Silver, D., Sugiyama, M.,
  Uchibe, E., and Morimoto, J.}
\newblock Deep learning, reinforcement learning, and world models.
\newblock {\em Neural Networks\/} (2022).

\bibitem{mchoskey1999relativism}
{\sc McHoskey, J.~W., Betris, T., Worzel, W., Szyarto, C., Kelly, K., Eggert,
  T., Miley, J., Suggs, T., Tesler, A., and Gainey, N.}
\newblock Relativism, nihilism, and quest.
\newblock {\em Journal of Social Behavior and Personality 14}, 3 (1999), 445.

\bibitem{puterman1990markov}
{\sc Puterman, M.~L.}
\newblock Markov decision processes.
\newblock {\em Handbooks in operations research and management science 2\/}
  (1990), 331--434.

\bibitem{puterman2014markov}
{\sc Puterman, M.~L.}
\newblock {\em Markov decision processes: discrete stochastic dynamic
  programming}.
\newblock John Wiley \& Sons, 2014.

\bibitem{relativism2008stanford}
{\sc Relativism, M.}
\newblock Stanford encyclopedia of philosophy.
\newblock {\em URL: http://plato. stanford. edu/entries/moral-relativism/\/}
  (2008).

\bibitem{schultz2016dopamine}
{\sc Schultz, W.}
\newblock Dopamine reward prediction-error signalling: a two-component
  response.
\newblock {\em Nature reviews neuroscience 17}, 3 (2016), 183--195.

\bibitem{silver2017mastering}
{\sc Silver, D., Schrittwieser, J., Simonyan, K., Antonoglou, I., Huang, A.,
  Guez, A., Hubert, T., Baker, L., Lai, M., Bolton, A., et~al.}
\newblock Mastering the game of go without human knowledge.
\newblock {\em nature 550}, 7676 (2017), 354--359.

\bibitem{singer2011expanding}
{\sc Singer, P.}
\newblock {\em The expanding circle: Ethics, evolution, and moral progress}.
\newblock Princeton University Press, 2011.

\bibitem{sloman2005problem}
{\sc Sloman, S.~A., and Lagnado, D.~A.}
\newblock The problem of induction, the cambridge handbook of thinking and
  reasoning (kj holyoak and rg morrison, editors), 2005.

\bibitem{sutton2018reinforcement}
{\sc Sutton, R.~S., and Barto, A.~G.}
\newblock {\em Reinforcement learning: An introduction}.
\newblock MIT press, 2018.

\bibitem{sutton1999policy}
{\sc Sutton, R.~S., McAllester, D., Singh, S., and Mansour, Y.}
\newblock Policy gradient methods for reinforcement learning with function
  approximation.
\newblock {\em Advances in neural information processing systems 12\/} (1999).

\bibitem{volkow2015brain}
{\sc Volkow, N.~D., and Morales, M.}
\newblock The brain on drugs: from reward to addiction.
\newblock {\em Cell 162}, 4 (2015), 712--725.

\bibitem{watkins1992q}
{\sc Watkins, C.~J., and Dayan, P.}
\newblock Q-learning.
\newblock {\em Machine learning 8\/} (1992), 279--292.

\bibitem{wise2004dopamine}
{\sc Wise, R.~A.}
\newblock Dopamine, learning and motivation.
\newblock {\em Nature reviews neuroscience 5}, 6 (2004), 483--494.

\bibitem{young2011collective}
{\sc Young, G., and Young, G.}
\newblock Collective intelligence and multilevel selection.
\newblock {\em Development and Causality: Neo-Piagetian Perspectives\/} (2011),
  733--758.

\end{thebibliography}
\bibliographystyle{acm}

\end{document}